\documentstyle[preprint,eqsecnum,aps]{revtex}
\begin{document}
\draft
\title{
Higher Order Correlations in Quantum Chaotic Spectra}
\author{Pragya Shukla$^{*}$} 
\address{Centre Emile Borel, Institut Henri Poincare
Universite Pierre et Marie Curie, Paris, France\\
and\\
Fachberich Physik, Universitat-GHS Essen, D-45117, Essen, Germany.}
\maketitle
\begin{abstract}
	The statistical properties of the quantum chaotic spectra
	have been studied, so far, only up to the second
	order correlation effects. The numerical as well as the 
	analytical evidence that random matrix theory can successfully
	model the spectral fluctuatations of these systems is available 
	only up to this order. For a complete understanding of spectral 
	properties it is highly desirable to study the higher
	order  spectral correlations. This will also inform us about the 
	limitations of random matrix theory  in modelling  the properties 
	of quantum chaotic systems. Our main purpose in this
	paper is to carry out this study by a semiclassical
	calculation for the quantum maps; however results are also valid
	for time-independent systems. 
\end{abstract}
\pacs{	PACS numbers: 05.45+b, 03.65 sq, 05.40+j}
\section{Introduction}
\label{int}
		In generic hamiltonian systems with many degrees of freedom, 
	the classical dynamics shows an enormous richness in structure, 
	increasing with the interaction between degrees of freedom. The 
	classical motion is mainly of two types, integrable and chaotic; 
	see [1] for details. This paper deals with the quantum properties 
 	of hamiltonians whose classical limit is chaotic.

		The strongly chaotic nature of underlying classical dynamics 
	intutively suggest us to expect some kind of random behaviour in
	quantum dynamics too. This is because the classical dynamics is
	indeed a limit ($\hbar=0$) of quantum dynamics and therefore the
	nature of former should be somehow reflected in the latter. In fact, 
	various analytical and numerical studies (see [2] and references
	there in) have confirmed that the 
	manifestation of chaotic behaviour in quantum dynamics occurs through 
	randomization (partial or full) of matrices of associated quantum 
	operators. The 
	spectral and strength fluctuations of these operators can be 
	well-modeled (up to second order correlations) 
	by one of the various universality 
	classes of random matrices. Most common among these are the 
	Gaussian orthogonal and unitary ensembles, GOE and GUE, the 
	analogous circular ensemble models COE and CUE [3].
	The former pertain to autonomous systems whereas the latter have
	application in the study of non-autonomous systems such as quantum
	maps.

		The presence of RMT type spectra in quantum chaotic systems
	can be explained by the Gutzwiller-semiclassical quantization
	scheme [4] for time-independent systems which uses the elegant 
	technique of path integral sum given
	by Feynman and relates the chaotic manifolds of classical dynamics 
	to the eigenfunctions of quantum dynamics. A similar formulation 
	is given for time-evolution operators of quantum maps too [5].
	The spectral fluctuation 
	measures can then be determined approximately by using the principle
	of uniformity [6] which is based on the uniform distribution of 
	periodic
	orbits at large time scales and gives a technique to evaluate the 
	sum of periodic orbit contributions. Using this technique for
	autonomous hamiltonians, Berry [7] provided an explicit expression for 
	the semiclassical form factor $K_{2}(\tau)$ -the fourier transform 
	of the $2$-level spectral correlation function-for values of $\tau$ 
	in the range $\tau<<1 $ (the time measured in units of $2\pi\hbar 
	\overline d$, 
	where $\overline d$ is the mean spectral density). This result has
	an exact analogy with the corresponding RMT behaviour; following 
	essentially the same technique as used by Berry for autonomous 
	hamiltonians, this analogy can be proved for quantum maps too [2]. 
	In 
	the region $\tau >> 1$ too, the limiting behaviour was analysed by 
	Berry using a semiclassical sum-rule which makes use of the properties
	of the function related to the quantum-mechanical density of states.

	 Notwithstanding the good agreement between RMT and statistical 
	quantum chaos up to $2^{nd}$ order correlations (long and very long), 
	still there is 
	no reason to believe that the Random Matrix Theory (RMT) can model 
	all $n^{th}$ order 
	spectral as well as strength correlations.
	 The numerical studies for many systems (e.g Baker map
	[8], Quantum Kicked Rotor [2], etc.) have already indicated
	that even second order correlation effects, when considered on
	short time scales (i.e very long range correlations) , do not follow 
	random matrix prediction and are non-universal. This is the range 
	where the classical dynamics is still diffusive and periodic orbits 
	are not yet uniformly distributed. The deviation from RMT in this 
	range agrees well with our intution as one should expect RMT to be 
	applicable only on those time-scales where the variables associated 
	with classical dynamics are random enough to fully randomize the 
	matrices associated with corresponding quantum operators.
	Moreover, the sum rules for the matrix elements of quantum
	chaotic operators [9] have already been found, differing from
	those of RMT. But a study of higher order correlations between 
	zeros of Reimann-zeta function shows a good agreement with RMT 
	[10,11]      

          	       Thus it is relevant to know as to what
       properties and up to what order the behaviour of
       quantum operators can be modeled by RMT
	and when it ultimately breaks down. Our attempt, in
       this paper, is to make a comparative study of one such property, 
       namely the  $n^{th}$ order spectral correlation as all $n$-level
	spectral fluctuation measures can be expressed in its terms. We fufill 
	this goal by carrying out a semiclassical study of the fourier
	 transform of $n$-level correlation function $R_n$; the reason to
	consider the fourier tranform lies in the covenience of its  
	analytical as well as numerical calculability. We proceed as follows. 

         	The Gutzwiller formulation gives us the density of states as
	a sum over periodic orbits and this gives rise to periodic orbit-
	interaction terms in $n$-level density correlation function. Berry, 
	in order 
	to obtain result for $2$-level form factor, neglected the 
	contribution from these interacting terms as a first order 
	approximation (the so called diagonal approximation). But, for 
	a complete evaluation of form factor, one has to calculate the 
	contribution due to interacting terms. The lack of the knowledge of
	action correlations handicaps us from doing so. One attempt in 
	this direction was made in ref.[12], in which by assuming the 
	complete validity of random matrix theory, the periodic orbit 
	 correlations
	were calculated from the RMT form factor. These when compared with 
	numerically obtained correlations (for Baker map, Hyperbola billiard 
	and perturbed Cat map) showed a good agreement. 
	The numerical study of these actions also indicated the presence 
	of an uncorrelated component  
	(exponentially larger than 
        correlated part); in this paper we use 
	this fact. 
        We assume that, on long time scales as a first order approximation, 
	actions are 
	uncorrelated and calculate the $n^{th}$ order form factor. Under 
	this approximation, the result 
        turns out to be same as that of RMT which is also confirmed from the 
	numerical analysis for at least two higher order fluctuation 
	measures, given in this paper. For 
        a complete calculation of $n^{th}$ order form factor, the action 
        correlations should also be taken into account which will 
	determine the 
        higher order terms. 

          	In this paper, we present our semiclassical study for 
	quantum maps but the method can easily be generalized for time-
	independent systems and one obtains the same results. Similarly 
        the RMT is given only for Circular ensembles but, once again, the
	final results are valid for Gaussian ensembles too which follows 
        due to GE-CE equivalence for large dimensions).      

		This paper is organized as follows:

	In section \ref{rmr}, we briefly review the definition of
	various random
	matrix ensembles. For later use, we also discuss the relation
	between the $n$-level form factor and correlation functions. The
	section\ref{qmc} deals with a brief review of the fundamentals
	of quantum
	maps and the earlier obtained results for $2$-level form factor.
	Both the section\ref{rmr} and section\ref{qmc} are included in
	this paper so as to
	clarify the ideas used in section\ref{hsc} which deals with the higher
	order correlations and form factos for the quantum spectra with
	exact symmetry. In section\ref{nv}, we numerically study the higher 
        order fluctuation measures, namely, skewness and excess for a 
        prototype quantum chaotic system, that is, Kicked Rotor and compare 
        them with those of RMT.  We summarize our results in 
	section\ref{con}.      
\section{ Preliminaries}
\label{pr}
\subsection{Random Matrix Results
\label{rmr}}
		We briefly outline here the results for $n^{th}$ order form
factor (i.e fourier transform of $n$-point  density correlation function)
for the eigenvalues of the equilibrium circular ensembles
of the random matrix theory.
\subsubsection{The Circular ensembles}
          	  The circular type equilibrium ensembles are ensembles of
   unitary matrices $U_\beta$ in which the distinct non-zero matrix
   elements of U
are distributed independently as zero-centered random variables; $\beta$
defines the number of independent components of the matrix elements of
U. There are three such ensembles, characterized 
by $\beta$, namely COE, CUE and CSE for $\beta$=1,2 and 4 respectively. These 
Universality classes are determined by the invariance of the system under 
time-reversal transformation (TR) (or more generally antiunitary 
transformation) 
and are described by the invariance of the ensemble measure: invariance under 
orthogonal or symplectic transformations for TR-invariant systems and under 
unitary transformations for TR-non-invariant ones. The invariance restricts 
the allowed space of matrices, for example, to that of symmetric unitary 
matrices for orthogonal invariance.      
\subsubsection{n-point Correlators}
The afore-mentioned unitarity of $U$ implies 
that its eigenvalues $exp(iE_j)$ lie on the unit circle in the complex plane, 
where exponents $E_j$'s are termed as eigen-angles. 
The density of states is then defined by
\begin{eqnarray}
\rho (E) & = & \sum_{j=1}^N \sum_{k=-\infty}^{\infty} 
\delta (E-2\pi k -E_j) \\  
& = & {1\over 2\pi} \sum_{n=-\infty}^{\infty} {\rm exp}(inE) 
Tr(U^n) 
\end{eqnarray}
and has the mean value $<\rho>=N/2\pi$. 

For analytical studies of the
spectrum, it is a usual practice to calculate the level density  
correlations. For cases where the level density $\rho(E)$ can be written as 
the sum of a smooth part $<\rho(E)>$ 
and a fluctuating component $\delta \rho(E)$, it is preferable to 
study the correlations $R_k$ between the fluctuating parts of the density. 
The $R_k$'s can be defined as follows,
\begin{eqnarray}  
R_k(E_1,..,E_k) &=& 
{{<\delta\rho(E_1) \delta\rho(E_2)....\delta\rho(E_k)>}\over
{<\rho(E_1>...<\rho(E_k)>}} 
\end{eqnarray}
here $\delta\rho(E)= \rho(E)-<\rho(E)>$, $E_j=E+l_jD$ for $j=1,2..,k-1$ 
and $E_k=E$ with $D$ as the mean spacing and $< >$ implying the averaging
over variable $E$ for ranges containing sufficient number of mean spacings.
By substracting $<\rho(E)>$ from eq.(2.1) and using notation 
$Tr(U^n)=t_n$, $\delta\rho(E)$ can further be written as follows,
\begin{eqnarray}
\delta\rho(E) &=& 
{1\over 2\pi} \sum_{n=-\infty,\not=0}^{\infty} t_n \: {\rm exp}(inE) 
\end{eqnarray}
The substitution of eq.(2.4) in eq.(2.3) gives   
\begin{eqnarray}  R_k(E_1,..,E_k) &=&
{1\over N^k} {\sum_{J} { <t_{j_1}t_{j_2}...t_{j_k}}>
\left\langle{\rm exp}[i(\sum_{m=1}^{k} j_m)E]\right\rangle_E  
{\rm exp}[iD(\sum_{m=1}^{k-1} j_ml_m)] } 
\end{eqnarray}
here $\sum_{J}$ implies the summation over all indices $j_1,j_2,..,j_k$ 
with each index varying from $-\infty$ to $\infty$ except zero (that is, none 
of the indices take value zero). The averaging over $E$ reduces the eq.(2.5) 
in following form,
\begin{eqnarray} 
R_k(E_1,..,E_k) & = &
{1\over N^k} {\sum_{J} {<t_{j_1}t_{j_2}...t_{j_k}}> 
\delta\left(\sum_{m=1}^k j_m \right) {\rm exp}[iD(\sum_{m=1}^{k-1} j_ml_m)] 
} \\ 
& = & {1\over N^k} \sum_{J}{'} 
{< t_{j_1}t_{j_2}...t_{-(\sum_{m=1}^{k-1}j_m)}}>
 {\rm exp}[iD(\sum_{m=1}^{k-1} j_ml_m)] \end{eqnarray}
here $\sum^{'}$ implies the $\sum_{J}$ subjected to condition that 
$\sum_{m=1}^{k-1} j_m \not=0$.      	 
        In semiclassical analysis,
 instead of dealing directly with $R_k$, it is easier to calculate the
$k^{th}$ order form
factor, defined as follows,
\begin{eqnarray} 
K_k (\tau_1 , ..,\tau_{k-1}) & = &
\int exp{\left[2\pi i \sum_{j=1}^{k-1} (E_j -E_k)\tau_j \right]} 
R_k (E_1 ,...,E_k) dE_1 ... dE_{k-1}  \\ 
& = &\int exp{\left[2\pi i \sum_{j=1}^{k-1} 
l_j \tau_j \right]}
R_k (l_1 ,...,l_{k-1}) dl_1 ... dl_{k-1} 
\end{eqnarray}
Substitution of eq.(2.5) in eq.(2.7) gives $K_k$ in terms of the traces,
\begin{eqnarray}
K_k &=& {1\over N^k} {\sum_{J}{'} 
{< t_{j_1}t_{j_2}...t_{-(\sum_{m=1}^{k-1}j_m)}}>
{\prod_{m=1}^{k-1} \delta(\tau_m+{j_m\over N})}} 
\end{eqnarray} 
As in this study we confine ourselves to calculation of $K_k$ only for 
$|\tau_m| < 1$, $m=1,2,..(k-1)$, for these values of $|\tau_m|$'s only 
those terms in the $\sum_J$ contribute which have indices 
$\{j_1,j_2,..,j_k\}$ much less than $N$. Therefore, 
the $\sum_J^{'}$ in eq.(2.5) can be replaced by $\sum_G^{'}$ in which 
the indices vary from some value $-n$ to $n$ where $n<N$ with all other 
conditions same.      

	The above result for $K_k$ can further be simplified by using the 
recently obtained result for the statistics of the traces [13] which 
indicates that the first few traces $t_1,t_2,..,t_n$ of large unitary 
matrices ($N$ large) taken from any of the circular ensembles display no 
noticeble correlation. The ensemble average of each of the traces vanishes 
(that is, $<{t_n}>=0$) 
for all the circular ensembles, due to uniformity of the distribution of 
$E_i$'s, the eigenangles; see ref.[13] for details. But note that 
$t_{n}$ and $t_{-n}$ are not independent of each other and 
one can show that[3,13]
\begin{eqnarray} < |t_n|^2> = \beta n/N \hspace{.3in} n<<N \end{eqnarray} 
where $\beta$ is $1$ or $2$ depending on whether the ensemble is COE or 
CUE respectively. Therefore, in general, 
$ {<\prod_{j=1}^{n} t_{j}}>=0$ if at least one $t_j$ 
 is such that its opposite $t_{-|j|}$ is not present in the product. 
This product exists only if the following condition is satisfied,
\begin{eqnarray} 
<{\prod_{j} t_j}>
& = & {< \prod_{l} |t_l|^2> } = \prod_{l} {< |t_l|^2>}
\end{eqnarray}
Now, as can be seen from eq.(2.10), for $k$-odd, every product  
appearing in the sum, contains odd number of $t_j$'s and therefore above 
condition can never be satisfied. This gives, for $k$-odd,
\begin{eqnarray}
K_{k-odd} (\tau_1 , \tau_2 ,...,\tau_{k-1}) &\simeq& 0, \hspace{.3in}
|\tau_i|_{i=1,..,k} < 1 \end{eqnarray}
on the other hand, the application of the condition (2.10) gives the 
following result 
for $k$ even,
\begin{eqnarray}
K_{k-even}(\tau_1 , \tau_2 ,...,\tau_{k-1}) &\simeq&
\alpha \sum_P \left[\delta(\tau_{p_1} + \tau_{p_2}) 
\delta(\tau_{p_3} + \tau_{p_4})....
\delta(\tau_{p_{k-3}} + \tau_{p_{k-2}}) \right] \nonumber \\
&  &{|\tau_{p_1} | |\tau_{p_3} | |\tau_{p_5}|...
|\tau _{p_{k-1}}| } \end{eqnarray}
where $\alpha=1$ for CUE and $2$ for COE. The $\sum_P$ implies the sum 
over all possible permutations of indices $p_1,p_2,...,p_{k-1}$ over the 
set ${1,2,...,k-1}$.      

	 The result given by eq.(2.13) and (2.14) are valid only when  
 each $|\tau|<1$. 
For cases with $|\tau|\simeq 1$ or $>1$, (i.e $n\simeq N$)  
one has to take into account the correlation between traces. Furthermore 
though the method adopted here for the derivation of $K_k$ result is 
applicable  
only for ensembles of unitary matrices but the final results are valid also 
for Gaussian ensembles. This follows due to the 
equivalence of fluctuation measures of the circular and Gaussian ensemble
in large dimensionality limit.      
\subsection { Quantum Map Vs Classical Map}
\label{qmc}
		A classical map can be described by a canonical mapping 
$M$ of coordiante variable $q$ and momenta variable $p$ at discrete 
time-step $t_n$ to those at $t_{n+1}$. 
\begin{eqnarray} 
\pmatrix {q_{n+1} \cr p_{n+1} \cr} = 
 M \pmatrix {q_{n} \cr p_{n} \cr} 
\end{eqnarray}
with $W(q_{n+1},q_n)$ as the generator of the map such that 
\begin{eqnarray} 
p_n = - { \partial W(q_{n+1},q_n) \over  \partial q_n}, \hspace{.3in}
p_{n+1} = {\partial W(q_{n+1},q_n) \over \partial q_{n+1}} 
\end{eqnarray}
The nature of the time-step considered can 
give rise to different kind of maps [1]. For example, for time-periodic 
hamiltonians, it is easier to study the dynamics in terms of fixed time steps 
(i.e the period of the hamiltonian), the related mapping known as 
stroboscopic mapping. For time-independent systems sometimes it is 
sufficient to consider only those steps of dynamics which occur
on a definite plane that is, intersections of trajectory with a
plane (instead of equal time-steps), known as Poincare mapping.        

	The quantization of a two-dimensional classical map when the phase 
space upon which it acts is compact leads to the construction of unitary 
matrices $U$ of a finite dimension $N$, and their semiclassical limit is 
obtained for $N\rightarrow \infty$.
For example, for a canonical mapping on a two-dimensional torus (here taken 
to be a two-dimensional phase space with periodicities $Q$ and $P$ 
 in $q$- and $p$-directions respetively), the 
corresponding quantum propagator acts in an $N$-dimensional Hilbert space 
and is represented by an $N\times N$ unitary matrix $U$. This follows  
as the number of states $N$ allowed to 
be associated, by quantization, with the finite classical space is restricted 
(Uncertainty Principle); $N$ is determined by the 
following relation,
\begin{eqnarray} 2\pi \hbar N &=& Q\times P  \end{eqnarray}
here $N$ plays the role of the inverse of Planck's constant, with 
$N \rightarrow \infty $ as semiclassical limit.       
\subsection {Semiclassical Form Factor for Quantum Maps : 
		Symmetry Preserving Cases}
\label{f2}
For quantum maps acting in a finite Hilbert space,  eqs.(2.3) and (2.9) 
can be used to write the quantum-mechanical two-level form factor 
   $K_2(\tau)$, 
\begin{eqnarray} 
K_2 (\tau) &=& {2\pi\over N^2} \int_0^1 {\rm d}r 
\quad \left\langle{\delta \rho\left(E+{r\pi\over N}\right)}
{\delta \rho\left(E-{r\pi\over N}\right)} 
{\rm exp}{\left({2\pi i r \tau}\right)}\right\rangle_E 
\end{eqnarray}
Using eq.(2.4) in eq.(2.18), $K_2 (k)$ can further be reduced in the 
following form,
\begin{eqnarray}
K_2 (\tau) & = & {1\over N} \sum_{n=0}^{N-1} \sum_{m=0}^{N-1} 
{\rm exp}[ik(E_n -E_m)] - N\delta_{n,0} \\
& = & {1\over N} |Tr (U^n)|^2 - N \delta_{n,0} \end{eqnarray}
Where $n=N\tau$. Now the 
semiclassical expression of $K_2 (\tau)$ can be obtained 
from above equation by using semiclassical form of $Tr(U^n)$ 
which can be expressed as a sum over periodic orbits in classical 
phase-space ($r=(q,p)$) [5],
\begin{eqnarray} 
Tr (U^n) &=& g\sum_j \sum_{m_j} A_j^{\{m_j\}} 
{\rm exp}{[i{m_j W_j\over \hbar} - i\pi \nu_j /2]}
\end{eqnarray}
here the amplitude 
$A_j$ (=$n_j |\partial(r_n -r_0)/\partial r_0|_{r_0 = r_n}^{-1/2} $
=$n_j ({\rm sinh} \alpha_j)^{-1}$)
of the contribution from each (multiply traversed) 
periodic orbit $j$, with period $n$, depends on the stability $\alpha_j$ 
of the orbit; for long periodic orbits $A_j$ can be approximated as 
$A_j\simeq n_j {\rm exp}(-\alpha_j)=n_j {\rm exp}(-\gamma  n_j)$ with 
$\gamma$ as entropy of the classical motion. 
$ W_j$ is the action for one traversal of the orbit, $m_j$ is the number of 
traversals, $n_j = n/m_j$ is the period of the orbit with single 
forward traversal and $\nu_j$ is the Maslov index.  
 The index $g$ refers to the number of symmetric analogs, existing in the 
phase-space, for the periodic orbit. While considering the long-range 
correlations which are mainly affected by the long periodic orbits, it is 
sufficient to consider $|m_j|=1$ (and therefore $n_j = n$); this follows 
due to the principle of uniformity which states that on large time scales 
periodic orbits tend to distribute uniformly in phase space, their density 
increasing exponentially while intensity decreasing. Thus, on large 
time-scales, long periodic orbits which are almost all primitive dominate 
the phase space.
      
The evaluation of  $|Tr (U^n)|^2$, in semiclassical limit
$N\rightarrow\infty$,  can be done as follows.
 As obvious from eq.(2.21), $|Tr(U^n)|^2$ contains terms of 
the type ${\rm exp}[i(W_j -W_k)/\hbar]$ and therefore, for a complete 
evaluation of $K_2 (\tau)$, it becomes important to study the distribution of 
amplitudes $A_j$ and actions $W_j$. But, in semiclassical limit
 $\hbar\rightarrow 0$, the significant 
contributions comes only from those orbit-interactions for which 
$W_j -W_i \leq o(\hbar)$. The contributions from other orbit-interactions 
 becomes negligible, in this limit, due to the presence of the  
rapid oscillations leading to destructive interferences. 
Thus, for the leading order semiclassical 
asymptotics of $|Tr(U^n)|^2$, one needs to consider only "diagonal" terms [7] 
with $W_i \simeq W_j$ which, in large-$n$ limit (such that $n/N = \tau<<1$) 
can be evaluated by invoking the Hannay's sum-rule for the amplitudes [6],
\begin{eqnarray} 
|Tr(U^n)|^2 \simeq g^2 \sum_j A_j ^2  
\end{eqnarray}
Now by using Hannay's sum rule for the intensities, which comes from 
the principle of uniformity [5] and is given by 
\begin{eqnarray} 
\sum^{ }_ iA^2_i \delta \left( \left\vert \tau \right\vert 
- {n_i \over N} \right)   &=&  {{N^2\left\vert \tau
\right\vert} \over g} \hspace{.3in} |\tau| << 1
\end{eqnarray}
one can obtain the two level form factor $K_2 (\tau)$ [7,2] 
 which turns out to be same as that for random
matrix ensembles, under small- $\tau$ approximation (eq.(2.14)), 
\begin{eqnarray} K_2 (\tau) &\simeq& g |\tau| \end{eqnarray}
Note that above result is valid only for $|\tau|<<1$ i.e $|n|<<N$. This limit 
of validity comes into existence due to considerations of only diagonal terms 
in the evaluation of $|Tr(U^n)|^2$. For cases $|\tau|\simeq 1$, one needs to 
consider the contributions due to the constructive interference of very long 
periodic orbits (with $W_i -W_j \simeq o(\hbar))$ too which once again 
requires an understanding of the distribution of periodic orbit actions. 
Moreover, in the derivation of spectral densty in terms of periodic orbits, 
the quasi-energy is assumed to be complex with a very small imaginary part 
$\epsilon$ (required to avoid the divergence of the formula, occuring 
for real energies). Due to the finiteness of $\epsilon$, 
the periodic orbits 
with period $n > n^*$ (that is the oscillations with energies $\delta E <
\epsilon$ where $\delta E=\hbar/n$ and $\epsilon =\hbar/n^*$)
can not be taken into account in this formulation.       

The result obtained in eq.(2.12) for semiclassical form factor is 
same as that for exact symmetry classes of RMT, under the same limit. For 
example, $g=2$ and $1$ give corresponding COE and CUE results respectively 
[2].      
\section{ HIGHER ORDER SPECTRAL CORRELATIONS}
\label{hsc}
	In this paper, we restrain ourselves to the study of the
   $ k $-level correlation functions in short time limits (where
   period $n$ of the longest periodic orbit, in phase space, is much
greater than unity but much less than $N$ i.e $ 1<<n<<N$ or $|\tau|=n/N<<1$
) which will allow us to use principle of uniformity in
the evaluation of cross multiplication of $k$ periodic orbit 
contributions, thus simplifying the calculations. For simplicity and to
   explain our method, we first calculate the $3^{rd}$ and 
$4^{th}$ order correlations and then generalize it to $k^{th}$ order 
correlations.
\subsection{ $3^{ {\rm rd}}$-Order Correlation}
For simplicity, let us first calculate the $ 3^{ {\rm rd}} $ order 
   correlation function. The substitution of eq.(2.4) and eq.(2.21)
   in eq.(2.3), with $k=3$, gives us
\begin{eqnarray} 
 R_3 \left(E_1,E_2,E_3 \right) & = & {g^3 \over N^ 3}
   \left\langle \sum^{ }_{ ijk} \sum^{ }_{ m_i,m_j,m_k=\pm
 1}A_iA_jA_k \right. \nonumber \\  
&  &\left\langle \left[ {\rm exp} \left(im_i \left({n_iE}  +
n_i\ell_ 1 {2\pi \over N}  + {W_i \over \hbar} \right) \right) {\rm exp}
\left(im_j \left({n_jE}  + n_j\ell_ 2{2\pi \over N}  + {W_j
\over \hbar} \right) \right) \right. \right. \nonumber \\  
&  & \left. \left. {\rm exp}
\left(im_k \left({n_kE}  + {W_k \over \hbar} \right) \right)
\right] \right\rangle 
\end{eqnarray}
Here $<>$ implies a local averaging with respect to $E$, that is, 
the energy averaging over ranges which are classically 
small but quantum mechanically large so that a large number of levels are 
included. For example, a good choice is to take the size of the averaging 
range to be $E$ itself i.e to define
\begin{eqnarray} 
<f(E)>_E  &=& {1\over E}\int_0^E f(E') {\rm d}E' \end{eqnarray} 
Therefore 
the variation of amplitude, in the above equation,  with respect to energy 
is very small (amplitude being a classical quantity)
 and can be ignored. This gives
\begin{eqnarray} 
R_3 \left(E_1,E_2,E_3 \right) & = & {g^3 \over N^ 3}
 \sum^{ }_{ ijk} \sum^{ }_{ m_i,m_j,m_k=\pm 1}A_iA_jA_k  
\quad {\rm exp}\left[\left(m_in_i\ell_1+m_jn_j\ell_2 \right)
{2\pi i\over N}\right]  \nonumber \\  
& &\left\langle {\rm exp}\left[\left(m_in_i+m_jn_j+m_kn_k \right)
{iE}\right]\right\rangle  
 \left\langle {\rm exp}\left[\left(m_iW_i+m_jW_j+m_kW_k \right)
{i \over \hbar}\right]\right\rangle
\end{eqnarray}
Due to averaging over E, the contribution of various terms,
in  eq.(3.3), will be
determined by the fact whether their exponents contain E or not; the terms 
containing a factor of type $ {\rm exp}[iEn]$ will not make 
any contribution. Thus we can 
divide all the terms into following two classes.

\vspace{.15in}
{\it Case (1): Terms with all $ n_i,n_j,n_k$ of the same sign
(i.e either all positive or all negative)}
\vspace{.15in}

        On averaging over $ E $, the contribution of these terms
   to $ R_3 $ turns out to be zero  due to presence of a factor of type 
$ {\rm exp} \left[\pm i \left( \left\vert n_i \right\vert + \left
\vert n_j \right\vert + \left\vert n_k \right\vert \right)E \right] $ 

\vspace{.15in}
{\it Case (2) 
 Terms with any two among $ \left(n_i,n_j,n_k \right) $ with same
 sign ($+$ or $-$) and third one with opposite sign.}
\vspace{.15in}
          
The terms under this case  contain a factor $ {\rm exp} \left[\pm
  \left(n_i+n_j-n_k \right)iE \right] $ 
 (and its permutations). As mentioned above,
 these terms will make a non zero contribution 
iff $ n_k=n_i+n_j $ (or $n_i=n_j+n_k, $ $ n_j=n_i+n_k). $
Thus eq.(3.3) can be reduced to the following form
\begin{eqnarray} 
R_3 \left(E_1,E_2,E_3 \right) & = & {g^3 \over N^ 3}
 \sum^{ }_{ ijk}A_iA_jA_k  \nonumber  \\  
&  &\sum_{m=\pm1} 
 {\rm exp} \left({2\pi m i \over N}\left(n_i\ell_ 1+n_j\ell_ 2 \right) 
 \right) <{\rm exp} \left({i m\over \hbar} \left(W_i+W_j-W_k \right)
 \right)> \nonumber \\
 &+&
 {\rm exp} \left({2\pi m i \over N}\left(n_i\ell_ 1-n_j\ell_ 2 \right) 
 \right)
 <{\rm exp} \left({i m\over \hbar} \left(W_i-W_j+W_k \right)
 \right)> \nonumber \\
 &+&
 {\rm exp} \left({2\pi m i \over N}\left(n_j\ell_2-n_i\ell_1 \right) 
 \right)
 <{\rm exp} \left({i m\over \hbar} \left(W_j-W_i+W_k \right)
 \right)>  \end{eqnarray}
with $ 2^{ {\rm nd}} $  and $3^{\rm rd}$ terms corresponding to 
$ n_i=n_j+n_k $ and $ n_j=n_i+n_k $ respectively.       

To evaluate terms of type $<{\rm e}^{i(W_i+W_j-W_k)/\hbar}>$ we 
   proceed as follows. Here $W_i$, the action of a periodic orbit
   with period $n_i$, can 
also be written as a sum of $n_i$ single step actions 
$ W_i = \sum_{l=0}^{n_i-1} W_l (q_{l+1},q_l)|_{q_{n_i}=q_0}$. For strongly 
chaotic dynanics and on large tine scales, these single step actions 
can be regarded as  
independent variables with a pair-correlation coefficient decaying  
exponentially to zero. An extension of central limit theorem therefore 
implies that $W_i$'s are Gaussian random variable on large
time-scales. Hence $\theta= W_i+W_j-W_k$ will also be a Gaussian
random variable with mean zero, the variance of which is given as follows,
\begin{eqnarray} Var \theta & = & <(W_i+W_j-W_k)^2> \\
&\simeq& <W_i^2> + <W_j^2> + <W_k^2> \simeq 3T \end{eqnarray} 
Here $T$ is the average time-period of periodic orbits given by 
$T=\hbar/\delta E$ where $\delta E$ is the energy range over which 
average is taken.

As on large time-scales, the phase-space is densely and uniforemly 
covered by periodic orbits and a typical trajectory can be approximated 
by a very long periodic orbit. This permits us to approximate the average 
of ${\rm exp}(iW_j)$ over all periodic orbits by a phase-space average. 
This gives $<{\rm exp}(i\theta)>={\rm exp}(-Var\theta)={\rm exp}(-3T)$.      
	Here, in eq.(3.6), the correlations between various actions i.e 
	terms of type $<W_iW_j>$ has been approximated to zero ($W_i$ 
assumed to be random variable). But, as mentioned in section\ref{int}, the 
correlation between actions are not entirely zero, that is, $W_i$'s are not 
exactly random variables. The distribution function $P(\theta)$ of these 
actions can be written as $P(\theta)=P_{random}+P_{correlated}$ where random 
part of the distribution $P_{random}$ dominates the non-random part 
$P_{correlated}$. Therefore, on large time scales $P(\theta)$ can be 
approximated  by a Gaussian which gives us first order term of $R_3$. To 
calculate higher order terms which are not negligible on very long time 
scales the correlations between actions must also be taken into
account.        

	To further simplify the calculation of $R_3$, 
$A_k (\simeq n_k {\rm e}^{-\gamma n_k})$, in eq.(3.4),  can be replaced by   
$A_iA_j(n_i^{-1}+n_j^{-1})$ for the terms which survive due to 
$ n_k=n_i+n_j $. Similarly for terms with $n_i=n_j+n_k $ 
or $(n_j=n_i+n_k) $, $A_k$ can be replaced by 
$A_i A_j^{-1} n_j (1-n_j n_i^{-1})$ and 
$A_j A_i^{-1} n_i (1-n_i n_j^{-1})$ respectively.
 This leads us to following form of $R_3$,
\begin{eqnarray}  R_3 & = & {g^3 \over N^3}  
\sum^{ }_{ ij}A_i^2 A_j^2 \left({1\over n_i}+{1\over n_j} \right)
\sum_{m=\pm 1}{\rm exp} \left({2\pi m i \over N} 
\left(n_i\ell_ 1+n_j\ell_ 2 \right) \right)  \nonumber \\  
&+& \sum^{ }_{ ij}  A^{2}_i n_j(1-n_j n_i^{-1})
\sum_{m=\pm 1}{\rm exp} \left({2\pi m i \over N} 
\left(n_i\ell_ 1-n_j\ell_ 2 \right) \right) \nonumber \\  
&+& \sum^{ }_{ ij} A^{2}_j n_i (1-n_i n_j^{-1})
\sum_{m=\pm 1}{\rm exp} \left({2\pi m i \over N} 
\left(-n_i\ell_ 1+n_j\ell_ 2 \right) \right)  
{\rm e}^{-3T}
\end{eqnarray}
here the $ 2^{\rm nd} $ term corresponds to $ n_i=n_j+n_k $ or $
n_k=n_i-n_j, $ the $ 3^{ {\rm rd}} $ term corresponds 
to $ n_j=n_i+n_k $ or $ n_k=n_j-n_i. $

The Fourier transform of $ R_3 $ gives us the $ 3^{ {\rm rd}} $
order form factor $ K_3 $, 
\begin{eqnarray}
 K_3 \left(\tau_ 1,\tau_ 2 \right) & =  & \int^{ }_{ } 
{\rm e}^{2\pi i \left[ \left(r_1-r_3 \right)\tau_ 1+ \left(r_2-r_3
\right)\tau_ 2 \right]}R_3 \left(r_1,r_2,r_3 \right) {\rm d} r_1
{\rm d} r_2 {\rm d} r_3 \\   
&  =&  \int^{ }_{ } {\rm e}^{2\pi i \left[\ell_ 1\tau_ 1+\ell_ 2\tau_
2 \right]} 
R_3 \left(\ell_ 1,\ell_ 2 \right) {\rm d} \ell_ 1 {\rm d} \ell_ 2
\end{eqnarray}
(where $ r_1-r_3=\ell_ 1$ and $ r_2-r_3=\ell_ 2). $ The eq.(3.9)
follows from eq.(3.8) as $R_3$ depends only on differences $r_1-r_3$,
$r_2-r_3$.

	The further calculation of $K_3$ can be done by substituting eq.(3.7) 
in eq.(3.9) and by making use of following equalities (Appendix A)
which follow from principle of uniformity,
\begin{eqnarray}  
\sum^{ }_ j n_j^a\delta \left( \left\vert \tau_ 2 \right\vert  -
{n_j \over N} \right) &=&  
{<n^a>\over g} = {f(0,a)\over g} 
\end{eqnarray}
and
\begin{eqnarray}  \sum^{ }_ i {A^2_i \over n_i} \delta 
\left( \left\vert \tau_ 1 \right\vert 
- {n_i \over N} \right)  &=& {N\over g}\end{eqnarray}
The result obtained depends on whether $\tau$'s  
are greater or less than zero. This gives rise to following three 
possibilities.
\vspace{.25in}
{\it Case (1): both $\tau_1,\tau_2 > 0 $ or $\tau_1,\tau_2 < 0 $}
\vspace{.25in}
\begin{eqnarray} 
 K_3 & = & {g^3 \over N^ 3}\left(\sum^{ }_{ ij} 
\left(A^2_i {A^2_j \over n_j}+A^2_j {A^2_i \over n_i}\right) \delta \left(
\left\vert \tau_ 1 \right\vert  - {n_i \over N} \right) \delta \left(
\left\vert \tau_ 2 \right\vert  - {n_j \over N} \right) \right)
{\rm e}^{-3T} \\
& \simeq &  g \left( |\tau_1| + |\tau_2| \right){\rm e}^{-3T } 
\end{eqnarray} 
It is obvious from above equation that $K_3$ falls very rapidly to 
zero for large $T$-values, that is, for time-scales on which
sufficiently long periodic orbits exist in the phase space. On long 
time-scales, therefore this is similar to the RMT result (eq.(2.13)).
\vspace{.15in}   
{\it Case (2): 
$\tau_1 > 0, \tau_2 < 0 $ or $\tau_1 < 0, \tau_2 > 0 $}
\vspace{.15in}
\begin{eqnarray} 
K_3 & = & {g^3 \over N^ 3} \sum^{ }_{ ij} \left(A^2_j {n_i} +
A^2_i { n_j } -A^2_j n_i^2 n_j^{-1}-A^2_i n_j^2 n_i^{-1}\right) 
\delta \left( \left\vert
\tau_ 1 \right\vert  - {n_i \over N} \right) \delta 
\left( \left\vert \tau_2 \right\vert  - {n_j \over N} \right) 
{\rm e}^{-3T} \\ 
& \simeq &  g \left[ |\tau_ 1| f_1 (|\tau_ 2|) + |\tau_ 2| f_1(|\tau_ 1|) 
-f_2(|\tau_1|)-f_2(|\tau_2|) \right]
{\rm e}^{-3T}  \end{eqnarray} 
On substituting values of $f_1=f(0,1)$ and $f_2=f(0,2)$ (Appendix A) in 
above equation, we get 
$K_3 \simeq g(|\tau_1|-|\tau_2|)({\rm e}^{N\tau_2}-{\rm e}^{N\tau_1})
{\rm e}^{-3T}$ where $N|\tau_1|$ and $N|\tau_2|$ are of the sam order as 
that of $T$. This results into a nearly zero $K_3$ on large-time scales  
  which is again similar to RMT result.       

	Note the above-mentioned similarity between $K_3$ results for
 quantum maps and RMT has been shown here only for those time-scales at 
which principle of uniformity is well-applicable to the distribution of 
periodic orbits. No conclusion can be drawn about the short-time scales 
from the above analysis although the deviation of two-point 
fluctuation measures for quantum maps from those of RMT [2,7] 
suggests us to expect the same for higher orders too.      
\subsection{ $4^{ {\rm th}}$-Order Correlation}
 To calculate the $ 4^{ {\rm th}} $ order correlation function, We
  substitute eq.(2.4) and eq.(2.21) in eq.(2.3), with $k=4$. This gives us
\begin{eqnarray} R_4 \left(E_1,E_2,E_3,E_4 \right) & = & {g^4 \over N^ 4}
 \sum^{ }_{ ijk} \sum^{ }_{ m_i,m_j,m_k=\pm 1}A_iA_jA_kA_r
 \left\langle {\rm exp}\left[\left(m_in_i+m_jn_j+m_kn_k+m_rn_r\right)
{iE}\right]\right\rangle \nonumber \\
&  &{\rm exp}\left[\left(m_in_i\ell_1+m_jn_j\ell_2 +m_kn_k\ell_3\right)
{2\pi i\over N}\right]  \nonumber \\
&  &\left\langle {\rm exp}\left[\left(m_iW_i+m_jW_j+m_kW_k+m_rW_r\right)
{i \over \hbar}\right]\right\rangle \end{eqnarray}
Again the significant contributions to $R_4$ come from following four 
types of terms,
\vspace{.15in}
{\it Case (1): Terms where pairwise cancellation occurs i.e terms 
with  $ n_i=n_k,n_j=n_r$ and $ W_i=W_k,W_j=W_r$  
(and their permutations)}
\vspace{.15in}

		  The contribution $R_{4i}$ from such terms 
can be written as follows,
\begin{eqnarray}  R_{4i} & = & {g^4 \over N^ 4} 
\sum_{perm} \sum^{ }_{ ij}A_i^2 A_j^2
\sum_{m=\pm 1} {\rm exp} \left({2\pi m i \over N} 
\left(n_i(\ell_ 1-\ell_3)+n_j\ell_ 2 \right) \right)  \end{eqnarray}
here $\sum_{perm}$ refers to the sum over all possible permutations of pairs.

\vspace{.15in}
{\it Case (2): 
 Terms with  $ n_i-n_j-n_k-n_r=0 $ (and other such permutations)}  
\vspace{.15in}     		 

	 The contributions to $R_4$ from terms with  
 $ n_i-n_j-n_k-n_r=0 $ can be written as follows,

\begin{eqnarray} & = & {g^4 \over N^ 4}  
 \sum^{ }_{ ijk}A_i^2 
\left({n_j n_k}-{n_j^2 n_k\over n_i} -{ n_j n_k^2\over n_i}\right) 
\sum_{m=\pm 1} {\rm exp} \left({2\pi m i \over N} 
\left(n_i\ell_ 1-n_j\ell_ 2-n_k\ell_3\right) \right) \nonumber \\ 
&  &\left\langle{\rm exp} \left( { i m\over
\hbar}\left(W_i-W_j-W_k-W_r\right) \right)\right\rangle 
\end{eqnarray}
Similarly one can write the contributions from terms with  
$ n_j-n_i-n_k-n_r=0 $ and $ n_k-n_i-n_j-n_r=0 $).  The 
symbol $R_{4ii}$ will refer to the sum of contibutions of all such terms.

The contribution  $R_{4iii}$ from term with $ n_r-n_i-n_j-n_k=0 $ is
\begin{eqnarray}  R_{4iii}& = & {g^4 \over N^ 4} 
\sum^{ }_{ ij}A_i^2 A_j^2 A_k^2
\left({1\over n_i n_j}+{1\over n_j n_k} -{1\over n_i n_k}\right) 
\sum_{m=\pm 1} {\rm exp} \left({2\pi m i \over N} 
\left(n_i\ell_ 1+n_j\ell_2 +n_k\ell_3\right) \right) \nonumber \\ 
&  & \left\langle{\rm exp} \left(- { i m\over
\hbar}\left(W_r-W_i-W_j-W_k\right) \right)\right\rangle \end{eqnarray}

\vspace{.15in}
{\it Case (3): Terms with  $ n_i+n_j-n_k-n_r=0 $ 
(and other such permutations)} 
\vspace{.15in}

		  The contribution to $R_4$ from terms with  
$ n_i+n_j-n_k-n_r=0 $ can be written as follows,
\begin{eqnarray}  & = & {g^4 \over N^ 4}  
\sum^{ }_{ ijk}A_i^2 A_j^2 n_k 
\left({1\over n_j}+{1\over n_i } -{n_k\over n_i n_j}\right) 
\sum_{m=\pm 1} {\rm exp} \left({2\pi m i \over N} 
\left(n_i\ell_ 1+n_j\ell_2 -n_k\ell_3\right) \right) \nonumber \\ 
& & \left\langle{\rm exp} \left(- { i m\over
\hbar}\left(W_i+W_j-W_k-W_r\right) \right)\right\rangle
\end{eqnarray}
Similarly one can write the contributions from terms with  
from terms $ n_i-n_i-n_k+n_r=0 $ and $ n_i-n_j+n_k-n_r=0 $.
The symbol $R_{4iv}$ refers to the sum of contibutions of all such terms.
      
Thus $R_4$ can be written as follows 
\begin{eqnarray} 
R_4  & = & R_{4i}+R_{4ii}+R_{4iii}+R_{4iv}  \end{eqnarray}
where in each of the contributions $R_{4i}...R_{4iv}$ the terms of type 
$<{\rm e}^{W_i+W_j+W_k-W_r}>$ can be replaced by ${\rm e}^{-4T}$ (as done 
earlier for $K_3$).       
The Fourier transform of $ R_4 $ gives us the $ 4^{ {\rm rd}} $
order form factor $ K_4 $, 
\begin{eqnarray} 
 K_4 \left(\tau_ 1,\tau_ 2,\tau_3 \right) & = &  (2\pi)^3\int^{ }_{ } 
{\rm e}^{2\pi i \left[\ell_ 1\tau_ 1+\ell_ 2\tau_2 +\ell_3\tau_3\right]} 
R_4 \left(\ell_ 1,\ell_ 2,\ell_3 \right) {\rm d} \ell_ 1 {\rm d} \ell_ 2 
{\rm d} \ell_ 3 \\
& = & K_i + K_{ii} + K_{iii}+K_{iv}  
\end{eqnarray}
where $K_{i,ii,iii,iv}= \int^{ }_{ } 
{\rm e}^{2\pi i \left[\ell_ 1\tau_ 1+\ell_ 2\tau_2 +\ell_3\tau_3\right]} 
R_{4(i,ii,iii,iv)} \left(\ell_ 1,\ell_ 2,\ell_3 \right) 
{\rm d} \ell_ 1 {\rm d} \ell_ 2 
{\rm d} \ell_ 3 $.

          Now $K_4$ can be calculated by substituting 
eqs.(3.17-3.20) in eq.(3.23) and using equalities eqs.(3.10,3.11). Again as 
   for $K_3$, the result depends on whether $\tau$'s are greater or
   less than zero. This gives rise to following three possibilities.

\vspace{.15in}
{\it Case (1)  $\tau_1,\tau_2,\tau_3 > 0 $ or 
  		$\tau_1,\tau_2,\tau_3 < 0 $}
\vspace{.15in}

	In this case, except for $K_{4iii}$, the contributions from all 
others, namely, $K_{4i}$, $K_{4ii}$ and $K_{4iv}$ are zero. Thus
\begin{eqnarray} K_4 = K_{4iii} & = & {g^4 \over N^ 4}
 \sum^{ }_{ ijk}\sum_{m=\pm 1} A^2_i A^2_j A^2_k \left( 
{1\over n_j n_k}+{1\over n_j n_k}+{1\over n_j n_k}\right)
\delta \left(\tau_ 1 - {mn_i \over N} \right) \nonumber \\ 
&  & \delta \left(\tau_ 2 - {mn_j \over N} \right) 
\delta \left(\tau_ 3 - {mn_k \over N} \right) 
{\rm e}^{-4T} \\
& \simeq & g\left(|\tau_1|+|\tau_2|+|\tau_3|\right){\rm e}^{-4T}
\end{eqnarray} 
Due to presence of the exponentially decaying factor, $K_4$ 
turns out to be approximately zero for large T-values which is again 
similar to RMT results. 

\vspace{.15in}
{\it Case (2)  Any two of $\tau_1,\tau_2,\tau_3 $ positive (negative)  
		and the third one negative (positive) }
\vspace{.15in}

		Let $\tau_i,\tau_j >0$ and  $\tau_k < 0 $ (where 
$i,j,k$ can take any of the values $1,2,3$). In this case,

\begin{eqnarray}  K_{4i} &\simeq & 
g\left[\delta\left(\tau_i+\tau_k\right)|\tau_i||\tau_j| 
+\delta\left(\tau_j+\tau_k\right)|\tau_i||\tau_j| \right] \\  
K_{4ii} & = & g
\left[ |\tau_ k| f_1 (|\tau_ i|) f_1(|\tau_j|) - f_1(|\tau_ j|) 
f_2(|\tau_i|)-f_1(|\tau_i|) f_2(|\tau_j|) \right]
{\rm e}^{-4T } \\  
K_{4iii} & = & 0 \\
K_{4iv} & = & g \left[|\tau_ i|f_1 (|\tau_ k|)  + |\tau_ j| 
f_1(|\tau_k|) -f_2(|\tau_k|)  \right]{\rm e}^{-4T } 
\end{eqnarray} 
As obvious from above equations, only $K_{4i}$ does not contain an 
exponentially decaying factor and therefore makes a non-zero contribution 
to $K_4$ in large time-limits.
The above results are valid also if $\tau_i,\tau_j <0$ and  $\tau_k > 0 $. 
A comparison with RMT results shows that this lowest order
contribution 
to $ K_4 $ is same as that in RMT for $ \left\vert \tau_ 1 \right\vert ,
\left\vert \tau_ 2 \right\vert , \left\vert \tau_ 3 \right\vert \ll 1. $ 
\subsection{ $k^{ {\rm th}}$-Order Correlation}
	The method used in calculation of $3^{rd}$ and $4^{th}$ order 
form factors can further be generalized to 
the $k^{th}$ order correlation function 
The substitution of eqs.(2.4) and (2.21) in eq.(2.3) gives us
\begin{eqnarray} R_k  & = & {g^k \over N^k} 
\sum^{ }_{ i_1,..,i_k} A_{i_1} A_{i_2}.... A_{i_k} 
\sum^{}_{ m_{1},..,m_{k}=\pm 1} \nonumber \\ 
&   & \left\langle {\rm exp} \left[im_{1} \left(n_{i_1}E
  + {2\pi \over N}  n_{i_1}\ell_ 1+{W_{i_1} \over \hbar} \right)
\right].... {\rm exp}
\left[im_{k} \left({n_{i_k}E}  + 
{W_{i_k} \over \hbar} \right) \right]\right\rangle \end{eqnarray}

	It can further be rearranged as follows,
\begin{eqnarray} 
R_k  & = & {g^k \over N^k} \sum^{ }_{ i_1,..,i_k} \prod_{l=1}^k A_{i_l}  
\sum^{}_{ m_{1},..,m_{k}=\pm 1} \left\langle {\rm exp}
\left[{iE} \sum_{l=1}^k m_{l} n_{i_l}\right]\right\rangle \nonumber \\ 
&  &{\rm exp} \left[{2i\pi \over N}\sum_{l=1}^{k-1} m_{l} n_{i_l}\ell_l
\right] \left\langle{\rm exp} \left[{i\over\hbar}\sum_{l=1}^k
m_{l} W_{i_l}\right]\right\rangle \end{eqnarray}
   
	As due to $\pm 1$ values taken by each $m_{l}$, $l=1,2,..,k$ there 
can be $2^k$ different combinations of $n_{i_l}$'s in the $1^{st}$ 
exponent of eq.(3.31). Let   
 $M(r)$ be the set of a particular choice of values for each $m_{l}$ in 
the set $\{m_{1}, m_{2},..m_{k}\}$    
 Therefore there can exist  $2^{k}$ such sets, denoted 
by $M(r)$ with $r=1\rightarrow 2^k$, of which only $2^{k-1}$ sets 
are distinct. Here two sets $M(r)$ and $M(r')$ are considered indistinct 
if the values of each $m_{l}$ in $M(r)$ is oppsite (in sign) to that 
in $M(r')$. Now eq.(3.31) can be rewritten as    
\begin{eqnarray} R_k &=& {g^k \over N^k}\sum^{2^k}_{r=1} F_{M_r} 
\end{eqnarray}
where $F_{M_r}$ is given as follows,
\begin{eqnarray} F_{M_r}  & = & 
\sum^{ }_{ i_1,..,i_k} {\prod_{l=1}^k A_{i_l}} 
 \left\langle {\rm exp}
\left[{iE}\sum_{l=1}^k m_{l} n_{i_l}
\right] \right\rangle \nonumber \\ 
&  &{\rm exp} \left[i{2\pi \over N}\sum_{l=1}^{k-1} m_{l} n_{i_l}\ell_l
\right] \left\langle
{\rm exp} \left[{i\over\hbar}{\sum_{l=1}^k m_{l} W_{i_l}} \right]
\right\rangle \end{eqnarray}
here, in above equation, the values taken by $m_l$'s  are the same as in
 $M(r)$.      

Now for the same reason, as given for $ 3^{ {\rm rd}} $ order correlation
function, 
 only those terms of eq.(3.33) will contribute significantly to $R_k$ for 
which the multiplying factor of $E$ in the exponent is zero. Due to 
$k$-summations over periodic orbits, each summation containing a large 
number of them, there are many possibilities, resulting in a zero coefficient
of $E$. These various possibilities may arise, due to the "group-wise
 cancellation of periods" in the $1^{\rm st}$ exponent of eq.(3.32),
 containing various groups of periods in the exponent,
 where in each group the positive traversals of 
few orbits are cancelled by the negative traversals of few other orbits.
Let $G$ stand for any division of indices ${1,2,..,k}$ into 
$q$ subgroups $(G_1,G_2,..,G_q)$ 
then a term appearing in eq.(3.33) will make a non-zero contribution 
to $F_{M_r}$ if it satisfies following condition,
\begin{eqnarray}
\sum_{G_j} m_{l} n_{i_l} &=& 0  \hspace{.3in} (j=1,2,..q)  \end{eqnarray}
where the summation is over indices $l$ present in 
subgroup $G_j$ and so on. As obvious, one of these subgroups will 
contain index $i_k$. Later we will need to distinguish it form other 
subgroups; let us call it $G_j'$. 
Thus $F_{M_r}$ can be rewritten as 
\begin{eqnarray} F_{M_r} & = & 
\sum_{G}\left\langle {\rm exp} \left[{i\over\hbar}
\sum_{l=1}^{p} m_{l} W_{j_l} \right]\right\rangle 
\prod_{j=1}^{q} C_{G_j} \end{eqnarray}
here $\sum_G$ implies the summation over all possible 
divisions of indices $i_1,i_2,..,i_k$ into various subgroups,
and $\prod_{j}$ implies the product of the contributions $C_{G_j}$  
from all ${G_j}$'s for one such division, where  
\begin{eqnarray} C_{G_j} & = &  
\sum^{ }_{j_1,..,j_k}\left(\prod_{l=1}^p A_{j_l} \right)  
\delta(\sum_{l=1}^{p} m_{l} n_{j_l}) 
{\rm exp} \left[{2i\pi \over N}\sum_{l=1}^{p} m_{l}n_{j_l}b_l
\right]  \end{eqnarray}
here the $p$ indices contained in subgroup $G_j$ are denoted by 
${j_1,j_2,..,j_p}$ with $m_l$ as the coefficient of $n_{j_l}$               
and $b_l$'s refer to $\ell$'s associated with these 
indices (e.g if $j_l=i_1$ then $b_l=\ell_1$ and so on); if $j_l=i_k$ then 
$b_l=b_k=0$. Note that, in the group $G_j'$, the index $j_1$ is always 
chosen to be $i_k$ (so as to simplify the presentation) with $b_1=b_k=0$.
Now by using $n_{j_1}= \sum_{l=2}^{p} \alpha_{l} n_{j_l}$, with 
$\alpha_l=-m_l/m_1$, and $A_{j_l}\simeq n_{j_l}{\rm e}^{-\gamma n_{j_l}} $,
one can show that 
\begin{eqnarray} 
\prod_{l=1}^p A_{j_l} &\simeq& 
\sum_{l=2}^p \alpha_l {A_{j_l}^{\alpha_l+1} \over n_{j_l}^{\alpha_l-1}} 
\prod_{s=2\not=l}^p  
{A_{j_s}^{\alpha_s+1} \over n_{j_s}^{\alpha_s}} \end{eqnarray} 
By using eq.(3.37), $C_{G_j}$ can further be reduced as follows 
\begin{eqnarray}
C_{G_j} & = & \sum^{ }_{j_1,..,j_l} \left(
\sum_{l=2}^p \alpha_l {A_{j_l}^{\alpha_l+1} \over n_{j_l}^{\alpha_l-1}} 
\prod_{s=2\not=l}^p  
{A_{j_s}^{\alpha_s+1} \over n_{j_s}^{\alpha_s}} \right) 
{\rm exp} \left[{2i\pi \over N}(\sum_{l=2}^{p}(b_l-b_1)m_l n_{j_l}) 
\right]  \end{eqnarray}
Note here that for $j=j'$, $b_1=0$.
It can be seen from above equations that the most significant 
contribution to $R_k$ comes from those 
terms where a pairwise cancellation of time-periods 
$ n_i,n_j,n_k,n_s,... $ (appearing as a factor of $ E $ in 
the exponent) as well as actions  $ W_i,W_j,W_k,W_s,... $ (so that there 
is no exponential decay) occurs which is possible only when $k$ is even. 
The contribution $R_{ki}$  from such terms 
can be written as follows,
\begin{eqnarray}
 R_{ki} & = & {g^k \over N^ k} 
\sum_{perm} \sum^{ }_{j_1,..,j_k }A_{j_1}^2
A_{j_3}^2....A_{j_{k-1}}^2 \nonumber \\
&  & \sum_{m=\pm 1} {\rm exp} \left[{2\pi m i \over N} 
\left(\sum_{l=1}^{(k-2)/2} n_{j_{2l-1}}
(b_{2l-1}-b_{2l})+n_{j_{k-1}}b_{k-1} \right) \right]  \end{eqnarray}
The $\sum_{perm}$ implies the summation over all possible 
permutations of indices $j_1,j_2,..,j_k$ taken from set
$i_1,i_2,..,i_k$.    
The contributions from all other terms contain an exponential term, 
with sum over actions (in units of $\hbar$), as its exponent  
(i.e term $<{\rm exp}(\sum m_{i_j}W_{i_j})>$. In large time limits, 
the application of CLT again permits us to replace this term 
by ${\rm e}^{-kT}$. The contribution of such terms to $R_k$ can
be given as follows 
\begin{eqnarray} 
R_{kii} & = & {g^k \over N^k}\sum_{r}\sum_{G}^{M_r}
 \prod_{j} \sum_{j_1,..,j_p}^{G_j} \left(
\sum_{l=2}^p \alpha_l {A_{j_l}^{\alpha_l+1} \over n_{j_l}^{\alpha_l-1}} 
\prod_{s=2\not=l}^p 
{A_{j_s}^{\alpha_s+1} \over n_{j_s}^{\alpha_s}} 
\right) \nonumber \\
&  & {\rm exp} \left[{2i\pi \over N}\left(
\sum_{l=2}^{p}(b_l-b_1)m_l n_{j_l} \right)\right] {\rm e}^{-kT} 
\end{eqnarray}
The substitution of eqs.(3.39) and (3.40) in eq.(2.9)  gives us
following result for $K_k$
\begin{eqnarray} K_k &=& K_{ki} + K_{kii}  \end{eqnarray}
where
\begin{eqnarray}  K_{ki} & = & {g^k \over N^ k} \sum_{perm} 
\sum^{ }_{j_1,..,j_k }A_{j_1}^2 A_{j_3}^2....A_{j_{k-1}} \nonumber \\
&  & \sum_{m=\pm 1} \delta(t_{2k-1}-{ m n_{j_{2k-1}}\over N})
\prod_{l=1}^{{k-1}\over 2}\delta(t_{2l-1}-{ m n_{j_{2l-1}}\over N})
\delta(t_{2l-1}+\tau_{2l}) \end{eqnarray}
and
\begin{eqnarray}
K_{kii} & = & {g^k \over N^k}\sum_{r}\sum_{G}^{M_r}
 \prod_{j} \sum_{j_1,..,j_p}^{G_j} \left(
\sum_{l=2}^p \alpha_l {A_{j_l}^{\alpha_l+1} \over n_{j_l}^{\alpha_l-1}} 
\prod_{s=2\not=l}^p {A_{j_s}^{\alpha_s+1} \over n_{j_s}^{\alpha_s}} 
\right) \nonumber \\
 &  &\left[\prod_{l=2}^{p}\delta(t_l+{m_l n_{j_l}\over N}) \right]
\delta(\sum_l^{G_j} t_l){\rm e}^{-k\tau} \end{eqnarray}
where $t_l$ is the $\tau$-variable associated with $b_l$ with $t_l=0$ if 
$b_l=\ell_k$. By using eq.(A3), one can further reduce above equations in  
following form,
\begin{eqnarray} K_{ki} &= & \sum^{ }_ P \left[\delta \left
(\tau_{p_1}+\tau_{p_2}\right)\delta
\left(\tau_{p_3}+\tau_{p_4} \right)...({k/2}-1) {\rm terms} \right]  
|\tau_{p_1}| |\tau_{p_3}| 
 ...  |\tau_{p_(k-1)}|\delta_{k,even} \end{eqnarray}
where $\sum_P$ refers to sum over all possible permutations of 
indices $\{p_1, p_2,..,p_4\}$ taken from set $\{1,2,..,k-1\}$,
and
\begin{eqnarray}  K_{kii} &=& {g} \sum_{r}\sum_{G}^{M(r)}
 \prod_{j} D_{G_j} {\rm e}^{-kT} \end{eqnarray}
with $D_{G_j}$ given as follows,
\begin{eqnarray} D_{G_j} & = & \delta(\sum_{l=1}^p t_l)
\left[\prod_{l=2}^p \delta(h_l+m_l)\right]
\sum_{l=2}^p \alpha_l f(\alpha_l+1,\alpha_l-1)
\prod_{s=2,\not=l}f(\alpha_s+1,\alpha_s) \hspace{.3in} 
{G_j\not\equiv G_j'}   \\
& = &\left[\prod_{l=2}^p \delta(h_l-m_l)\right]
\sum_{l=2}^p \alpha_l f(\alpha_l+1,\alpha_l-1)
\prod_{s=2,\not=l}f(\alpha_s+1,\alpha_s) \hspace{.3in} { G_j\equiv G_j'} 
\end{eqnarray} 
with $h_l=sign(t_l)$.
Here, as obvious, the 
contribution $K_{ki}$ exist only for for all even order form factors. 
For odd order form factors, only 
$K_{kii}$ contributes. But, on large time scales,the presence of an 
exponential decaying 
factor makes this contribution very small.
 Thus, on large time-scales (i.e $n\rightarrow \infty, N\rightarrow\infty, 
n/N <1$) one can write 
\begin{eqnarray} K_{k-odd} &\simeq& 0 \end{eqnarray}
A comparison of eqs.(3.44) and (3.47) with eqs.(2.14) and (2.13), 
informs us that 
the results obtained for both odd as well as even order form factors 
agree well with that of RMT.
\section {Numerical Verification}
\label{nv}
	This section contains a numerical study of $3^{rd}$ and $4^{th}$ 
 	order fluctuation measures, namely, skewness $\gamma_1$ 
	and excess $\gamma_2$. 

	 We choose 
	the kicked rotor system for this purpose as it 
   has been an active model of research, containing a 
   variety of features such as localization, resonance, dependence of
   the spectra on number theoretical properties etc and has been used as a
   model for a very wide range of physical systems. 
   For a better understanding, we briefly review the quantum and classical 
   mechanics of kicked rotor in this section. 
\subsection{The Kicked Rotor: Classical and Quantum Dynamics}
	   The kicked rotor can be described as a pendulum subjected to
periodic kicks (with period T) with the following Hamiltonian
\begin{eqnarray} H &=& {{\left(p+\gamma\right)^2}\over 2} + {K}\quad {\cos}
(\theta+\theta _0) \sum_{n=-\infty}^{\infty} \delta (t-nT)\end{eqnarray}
where $K$ is the stochasticity parameter. The parameter $\gamma$ and
$\theta_0$ are introduced in the Hamiltonian in order to mimick the
effects of the time-reversal (T) and the parity (P) symmetry breaking
in the quantum system.      

        The related quantum dynamics can be described, by using Floquet's 
theorem, by a discrete time evolution operator $U=BG$ where $B={\rm
exp}(-i K \cos(\theta+\theta_0)/ \hbar)$ and 
$G={\rm exp}(-i(p+\gamma)^2/4\hbar)$.
The nature of the quantum dynamics 
and therefore the statistical 
properties of the
associated quantum operators depend on $\hbar$ and $K$. For a rational value
of $\hbar T/2\pi$, the dynamics can be confined to a torus while for 
irrational value it takes place on a cylinder.  
 We employ
torus boundary conditions $(q'=q+2\pi, p'=p+2\pi M/T)$ by taking 
$\hbar T/2\pi=M/N$;
 both
$p$ and $\theta$ then 
have discrete
eigenvalues and $U$ can be reduced to a finite $N$-dimensional matrix of
the form [32-34]
\begin{eqnarray} U_{mn} & = & {1\over N}
{\rm exp}{\left[-i{K\over \hbar}cos{\left({2\pi m\over
N}+\theta_0 \right)}\right]} \nonumber \\
&  &\sum_{l=-N_1}^{N_1} {\rm exp}
\left[-i\left({\pi ^2 \hbar}l^2 - \pi \gamma l \right)\right] 
{\rm exp}\left[-i\left( {{l(m-n)}\over N}\right)\right]\end{eqnarray}
where $N_1 = (N-1)/2$ (with $N$ odd) and $m,n = -N_1, -N_1 +1,...,N_1$ .      

	The quantum dynamics has a time
reversal symmetry $T$ for $\gamma=0$ and a parity symmetry $P$ for $\theta_0
=0$. Though the $T$-symmetry may be violated for $\gamma\neq 0$, still a more
generalized antiunitary symmetry $S=TP=PT$ can be preserved in the system if
$\theta_0 =0$ [14,2]. For $K^2>>N\hbar$, the quantum dynamics is delocalized 
   in the phase space and two point spectral fluctuation measures
   have been shown to be well modeled by various symmetry classes
   of RMT [14,2].
In the opposite limit of weak chaos, namely, $K^2 << N\hbar$, the
eigenstates localize in the momentum space and one obtains a
Poisson distribution for the spectrum [2]. For numerical comparison of 
higher order measures with RMT, therefore, we choose various parameters 
such that condition $K^2 >>N\hbar$ is always satisfied.
\subsection { Numerical Study of Skewness and Excess} 
		In this section, we numerically study skewness and excess, 
that is, the $3^{rd}$ and $4^{th}$ order spectral fluctuation measure, 
for QKR-spectra and compare them with corresponding RMT results. 
   Although the $k^{th}$ order form factor can easily be calculated
   for quantum maps (see eq.(2.10)), we choose to study $\gamma_1$
   and $\gamma_2$ as corresponding numerical results (for form
   factors) in RMT are not available. But as both $\gamma_1$ and
   $\gamma_2$ can analytically be expressed in terms of $3^{rd}$
   and $4^{th}$-order correlations $R_3$ and $R_4$ (see ref.[15]
   for these expressions) and, therefore, are related to form
   factors $K_3$ and $K_4$ (eq.2.8), any conclusion about the 
   validity of RM model for the former measure will be applicable
   for the latter too.

Both skewness and excess are the functions of $3^{rd}$ and
$4^{th}$ order central moment of a distribution respectively
which contain the information about the probability of higher
order events as compared to the lower order. More precisely, the
skewness denotes the absence of symmetry in the distribution and
can be defined as follows, 
\begin{eqnarray} \gamma_1(r) &= &{\mu_3(r) \over \sigma^3(r)}
\end{eqnarray} 
          The excess $\gamma_2$ describes the difference between
   the kurtosis values (i.e the fourth central moment calculated
   in units of the square of second central moment) of the
   distribution and that of a normal distribution (kurtosis=3), 
\begin{eqnarray} \gamma_2(r) = {\mu_4(r) \over \sigma^4(r)}-3  
\end{eqnarray} 
  where $\sigma^2$ is the variance in the number of levels in a length
   of r mean spacings and $\mu_3$ and $\mu_4$ are corresponding
   $3^{rd}$ and $4^{th}$ central moments. If excess is
   less than zero, the curve is platykurtic and if it is positive, 
  leptokurtic.          

          For both $\gamma_1$ and $\gamma_2$ studies, the spectral 
   data consists of the eigenvalues of $50$ matrices of dimension
   $N=199$ obtained by diagonalizing $U$-matrix (eq.(4.2)) for 
various values of $K$ in the neighbourhood of $K\simeq 20000$. The choice 
   of such a high value of $K$ is made to ensure the
   delocalization of quantum dynamics which makes $U$ a full random
   matrix (see ref.[2]). 
Due to strong sensitivity of the eigenvalues to small changes in $K$, these 
sequences of quasienergies can be regarded as mutually independent.      

	The results obtained for $\gamma_1$ and $\gamma_2$ of the QKR-spectra 
   are displayed, as function of $r$ in figures $1$ and $2$;. For
comparison, the corresponding 
RMT-results (taken from ref.[15]) are also given in each of the figures. 
The good agreement between QKR and RMT results, indicated by each of these 
   figures reconfirms our semiclassical results obtained in 
section\ref{hsc}. Furthermore, as indicated by these figures, the higher order
   correlations e.g $3^{rd}$ and $4^{th}$ order, seem to be very
   weak, nearly zero, at very long energy ranges and levels seem to
   be uncorrelated while strong correlation seem to be existing for
   short ranges. A continuously decreasing value of $\gamma_1$
   implies the tendency of the distribution to appear more and more
symmetric as range of the distribution increases and finally
acquire the Gaussian form for long energy ranges. This conclusion
is also supported by our $\gamma_2$ study which shows a larger
probability of higher order events for QKR at small energy
ranges (as compared to Gaussian case) while, for large energy
ranges, it acquires the same form for both.
\section{Conclusion}
\label{con}
	We conclude this article with a summary of our principle results 
and a brief discussion of the open problems.       

	We have shown that, in long time limits, the higher order 
spectral correlations in quantum 
chaotic maps can be well modeled by corresponding ones in RMT.
 By using the example of kicked rotor, we 
   have verified this numerically too. These results are also valid for
   conservative Hamiltonians; this follows due to the similar
   expressions for the level density in terms of classical periodic
orbits for both the cases (see Ref.[4] for level density for
autonomous case). The various summation formulas used to
evaluate fluctuation measures such
as sum over amplitudes (Appendix A), 
whose derivation depends 
   on the uniform distribution of periodic orbits and therefore
   only on the strongly chaotic nature of dynamics, still remain
valid. Thus one obtains the similar results for form factors in autonomous
   systems. This should not be surprising, as both Gaussian
   ensembles (the ensembles of hermitian matrices and therefore of 
   conservative hamiltonians) and circular ensembles are known to
have the same statistical behaviour in semiclassical limit [16].

	Although the study presented here deals with higher order
correlations at fixed value of parameter, we expect the validity
of random matrix
model also for higher order parametric correlations in quantum
chaotic spectra, but again
only on long time-scales. The intuition, based on the analogy of
$2^{nd}$ order parametric density correlations for mesoscopic
systems with a disordered potential, one dimensional many body
hamiltonian (Calogero-Sutherland model), scattering systems, 
quantum chaotic systems and random matrix models, as well as the
existence of a common mathematical base (nonlinear
$\sigma$-model and supersymmetry
approach) further encourages us to hope the extension of this
analogy for higher order correlations too.   

	   The ignorance about action correlations 
on very long time scales handicaps us from doing the same analysis on 
these scales. It will also be of interest to study these correlations on 
   short time scales. It is on these scales where $2^{nd}$ order
spectral correlations deviate from RMT and show non-universal
behaviour. One expects to see similar deviations for higher orders too. 
For a complete understanding of higer order correlations in
quantum chaotic systems, therefore, 
one should study the action and periodic orbit correlations;
this will give us some insight in the behaviour of fluctuations
in other systems (above-mentioned) too.  We intend to do so in
future. 
\acknowledgements
I would like to thank A.Pandey for suggesting this study and 
J.P.Keating and Z.Rudnick for various suggestions.   
 I am grateful to 
CEB, Institut Henri Poincare, Paris and Fachberich Physik, Universitat-GHS 
   Essen, Germany too for support and environment in which this
   work could prosper. I also acknowledge financial support by the 
   Sonderforschungsbereich "Unordnung und gro$\beta$e Fluktuationen"
   of the Deutsche Forschungsgemeinschaft" during my stay in Essen.
\appendix
\section{Calculation of $f(a,b)$ }
	To evaluate the following sum \begin{eqnarray}  
f(a,b) & = & \sum^{ }_j {A_j^a\over n_j^b}\delta \left(\left\vert 
\tau \right\vert  - {n_j \over N} \right)  \end{eqnarray}
where $A_j$ is the amplitude and $n_j$ is the period of the orbit in the 
classical phase space.  We remind ourselves that, in the phase space, the 	
 orbits proliferate exponentially with time and the density 
of distribution of periods over long orbits is given by 
${\rm exp}(2\gamma |n|)/|n|$ with $\gamma$ as the entropy of the classical 
motion. In large time-limits. This, along with the approximation 
$A_j\simeq n_j exp(-\gamma |n_j|)$, enables us to make the following 
replacement 
\begin{eqnarray} 
\sum_j {A_j^a\over n_j^b} \delta(|\tau|-{n_j\over N}) & \simeq & 
   \sum_{n=0}^\infty {\rm exp}(-\gamma n(a-2)) n^{a-b-1}
   \delta(|\tau|- {n\over N}) \end{eqnarray}
which gives 
\begin{eqnarray}
f(a,b) &=&  {\rm exp}(-\gamma N|\tau|(a-1)) |\tau|^{a-b-1} N^{a-b} 
\end{eqnarray}

\begin{figure}
\caption{The behaviour of $\gamma_1 (r)$ with respect to $r$, 
with $N=199$, $\hbar=1$, $T=1$, $K=20002\rightarrow 20050$ and for       
 (a) $\gamma=0.0, \theta_0 = \pi/2N$, TR-preserved      
 (b) $\gamma=0.7071, \theta_0 = \pi/2N$, TR-broken      
The solid curve depicts the corresponding 
RMT behaviour, namely COE (in 1(a)) and CUE limit (in 1(b)).}       
\label{fig1}
\end{figure}
\begin{figure}
\caption{The behaviour of $\gamma_2 (r)$ with respect to $r$, 
with $N=199$, $\hbar=1$, $T=1$, $K=20002\rightarrow 20050$ and for       
 (a) $\gamma=0.0, \theta_0 = \pi/2N$, TR-preserved       
 (b) $\gamma=0.7071, \theta_0 = \pi/2N$, TR-broken      
The solid curve depict the corresponding 
RMT behaviour, namely COE (in 2(a)) and CUE limit (in 2(b)).}      
\label{fig2}
\end{figure}

\begin{references}
\bibitem[*] 
Present Address: Department of Physics, Condensed Matter Theory Unit, 
Indian Institute of Sciences, Banglore-560012, India;
E-Mail : Shukla@physics.iisc.ernet.in 
  
\bibitem{} 
A.M.Ozorio de Almeida {{\bf Hamiltonian Systems: Chaos and Quantization}} 
 (Cambridge Univ. Press, Cambridge 1988).

\bibitem{} 
P.Shukla and A.Pandey, Preprint;       
P.Shukla  Ph.D Thesis 1992 Jawahar Lal Nehru Univ., New Delhi.

\bibitem{} 
M.L.Mehta {{\bf Random Matrices}} (Acedemic Press, New York 1990).


\bibitem{} 
 M.C.Gutzwiller  J. Math. Phys. 12 343 (1971).  

\bibitem{} 
M.Tabor  Physica D6  195 (1983)  

\bibitem{} 
 J.H.Hannay and A.M.O.de Almeida  J. Phys. A 17  3429 (1984).

\bibitem{} 
M.V.Berry  Proc. R. Soc. A400  299 (1985).  

\bibitem{} 
P.W.O'Connor and S.Tomsovic Ann. Phys. 207 (1991). 

\bibitem{} 
 M.Wilkinson J. Phys. A 20 2415(1987) and 21  1173 (1988).  

\bibitem{} 
J.P.Keating and E.Bogomolny, Nonlinearity, (1995).

\bibitem{} 
Z.Rudnik and P.Sarnak, preprint.

\bibitem{}
N.Argmann, F-M.Dittes, E.Doron, J.P.Keating, A.Yu.Kitaev, M.Sieber and 
U.Smilansky, Phys.Rev.Lett., 71 4326 (1993).  

\bibitem{} 
F.Haake, M.Kus, H-J.Sommers,H.Schomerus and K.Zyczkowski, Preprint.

\bibitem{} 
Izrailev, Phys.Rev.Lett. 56 541 (1986).

\bibitem{} 
O.Bohigas, R.U.Haq and A.Pandey, Phys.Rev.Lett. 54 1645 (1985).      

\bibitem{}
A.Pandey and P.Shukla, J.Phys.A,

\end{references}
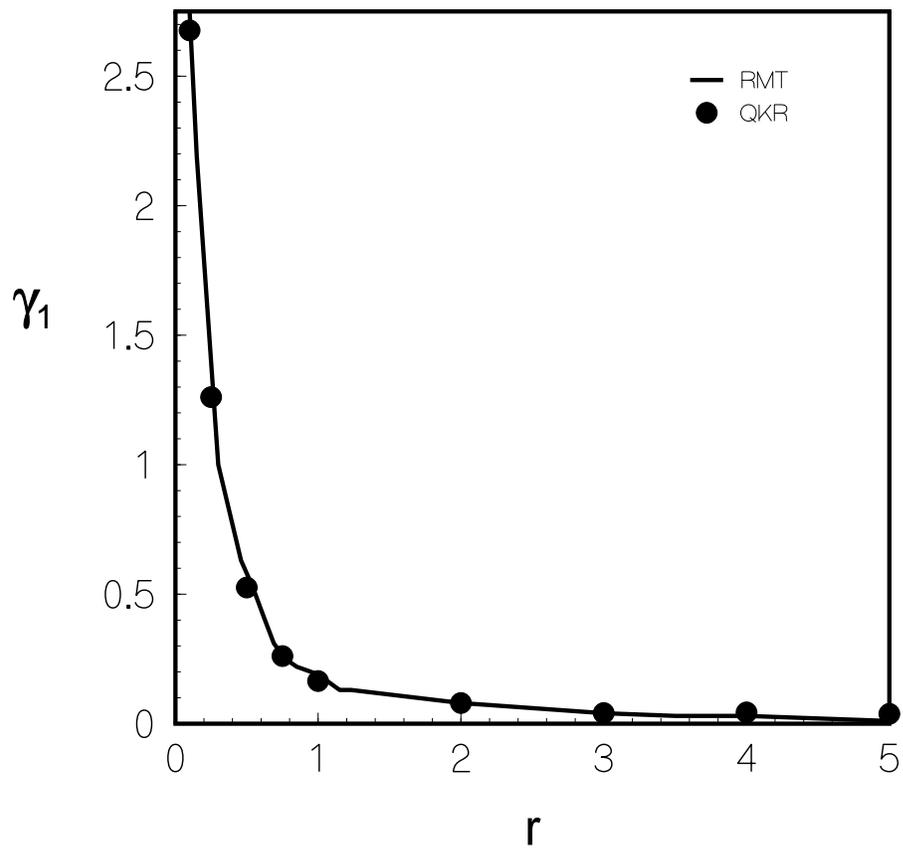
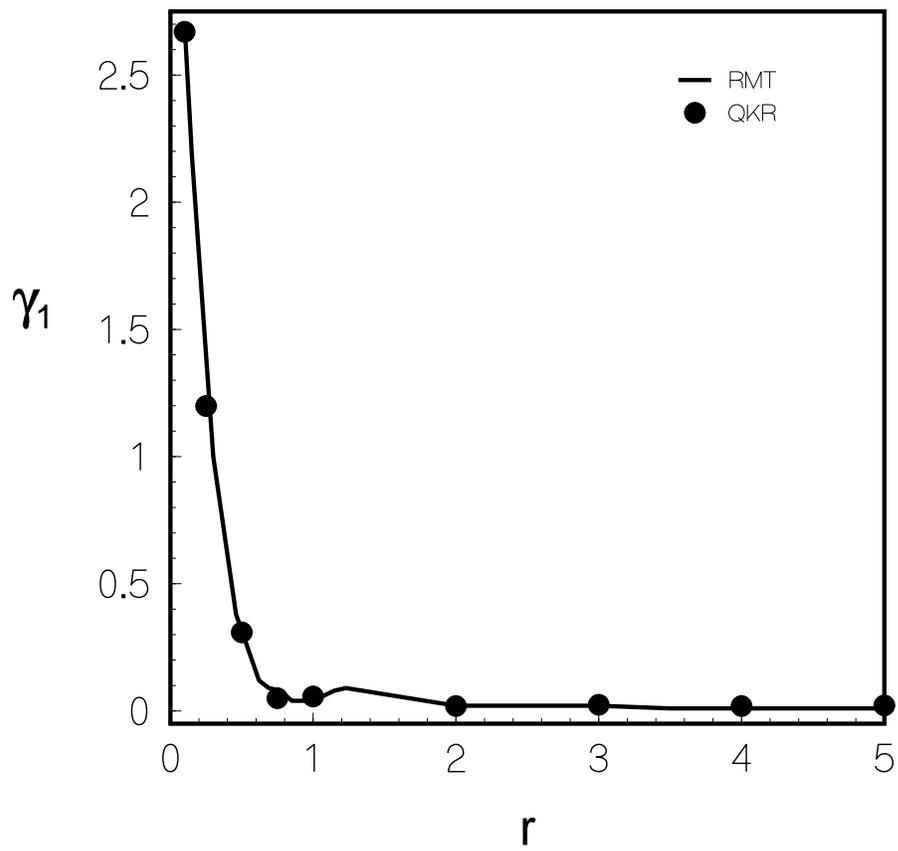
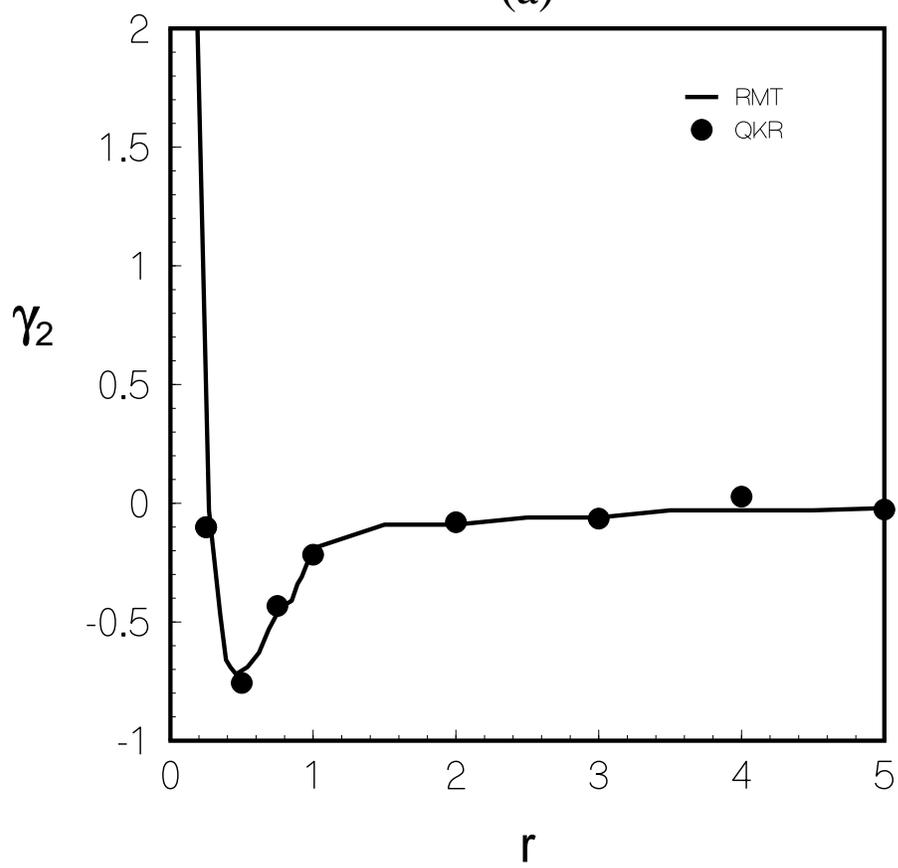
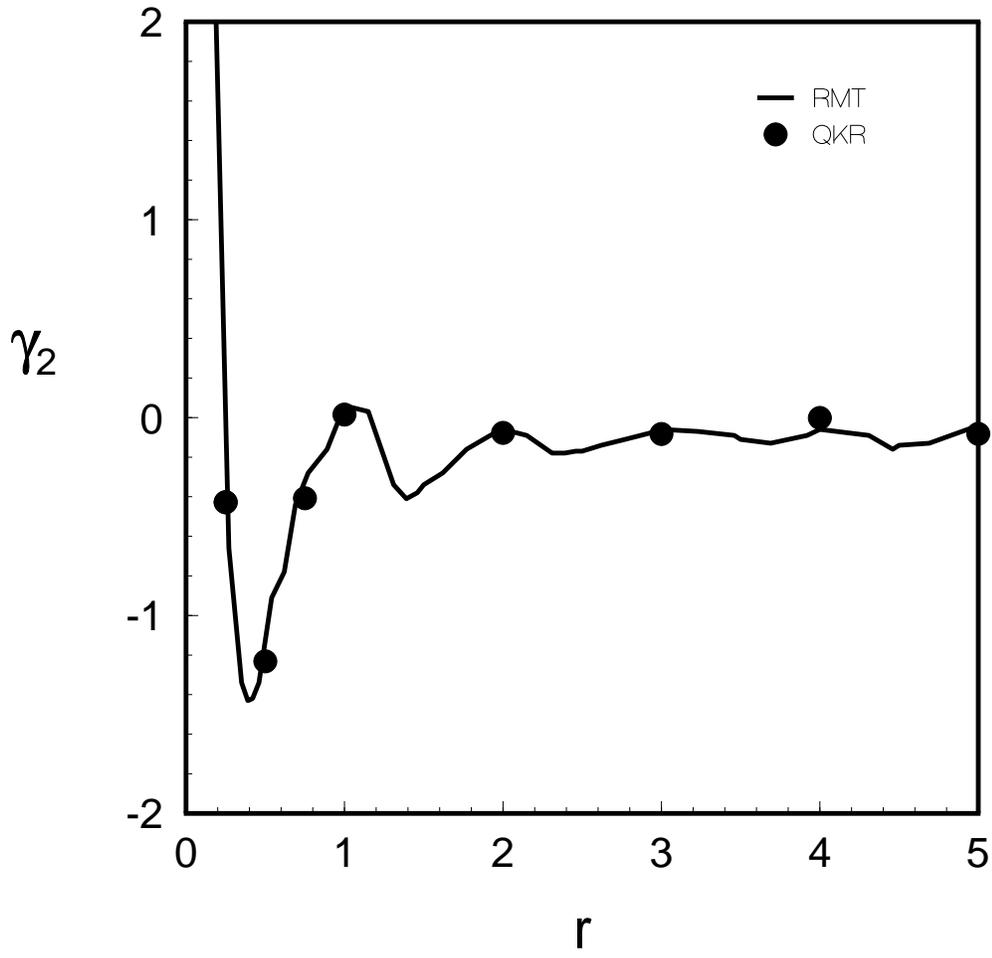
\end{document}